\documentclass[twocolumn,secnumarabic,amssymb, nobibnotes, aps, prd, showpacs, showkeys]{revtex4-1}

\usepackage{epsfig}

\begin{document}

\title{Inverse magnetocaloric effect in the spin-1/2 Fisher's
super-exchange antiferromagnet}%

\author{L. G\'{a}lisov\'{a}$^{1,}$}

\email[e-mail: ]{galisova.lucia@gmail.com}

\author{J. Stre\v{c}ka$^2$}%

\affiliation{$^1$Department of Applied Mathematics and Informatics,
             Faculty of Mechanical Engineering, Technical University,
             Letn\'{a} 9, 042 00 Ko\v{s}ice, Slovak Republic
             \\
             $^2$Institute of Physics,
             Faculty of Science, P.~J.~\v{S}af\'{a}rik University,
             Park Angelinum 9, 040 01 Ko\v{s}ice, Slovak Republic}%

\begin{abstract}
The isothermal entropy change and the adiabatic temperature change are rigorously calculated for the exactly solved spin-1/2 Fisher's super-exchange antiferromagnet  in order to examine magnetocaloric properties of the model in a vicinity of the second-order phase transition. It~is shown that the large inverse magnetocaloric effect occurs around the temperature interval \mbox{$T_c(h\neq0) < T < T_c(h = 0)$} for any magnetic-field change $\Delta h\!: 0\to h$. The most pronounced inverse magnetocaloric effect can be found for the magnetic-field change, which coincides with the critical field of a zero-temperature phase transition from the antiferromagnetically ordered ground state to the paramagnetic one.
\end{abstract}

\pacs{05.50.+q, 75.30.Et, 75.30.Sg, 75.30.Kz}

\keywords{Fisher's super-exchange antiferromagnet, inverse magnetocaloric effect, exact results}

\date{\today}%
\maketitle

\section{Introduction}
The magnetocaloric effect (MCE), which is characterized by an isothermal
change of the entropy or an adiabatic change of the temperature upon magnetic-field variation, enjoys a great scientific interest mainly because of its immense application potential~\cite{1}. Besides a  conventional MCE observed in regular ferromagnets or paramagnets, there may also be detected an inverse MCE, namely, in ferrimagnetic or antiferromagnetic materials. In the former case the system cools down when the magnetic field is removed adiabatically, while in the latter case it heats up.

However, the MCE has been so far rigorously studied only in one-dimensional spin systems (see e.g.~Ref.~\cite{2} and references therein) due to a lack of exactly solved spin models in higher dimensions accounting for a non-zero magnetic field. Theoretical description of the conventional and inverse MCE in two- and three-dimensional magnetic systems is thus usually based on some approximative method~\cite{3}.

The main goal of this work is to investigate the MCE in the exactly solved spin-1/2 Fisher's super-exchange antiferromagnet~\cite{4,5}, which allows an exact theoretical description of this phenomenon also in a  two-dimensional spin model.

\section{Fisher's super-exchange antiferromagnet}
\label{sec:model}

The spin-1/2 Fisher's super-exchange antiferromagnet represents the spin-1/2 Ising model on a decorated square lattice, in which the antiferromagnetic (ferromagnetic) coupling $J>0$ on horizontal (vertical) bonds are supposed together with the external magnetic field $h$ acting on decorating spins. The Hamiltonian of the model reads
\begin{equation}
H = J\sum_{\left\langle i, j\right\rangle}\mu_i^z\sigma_j^z -  J\sum_{\left\langle k, j\right\rangle}\mu_k^z\sigma_j^z - h\Big(\sum_{i}\mu_i^z +\sum_{k}\mu_k^z\Big).
\label{eq:H}
\end{equation}
In above, $\mu_{i(k)}^z = \pm1/2$ labels the decorated Ising spin at $i$th horizontal ($k$th vertical) bond and $\sigma_j^z=\pm1/2$ denotes the nodal Ising spin at $j$th site of the original square lattice. The first (second) summation in the Hamiltonian~(\ref{eq:H}) is thus carried out over nearest-neighbour lattice sites on the horizontal (vertical) bonds, while the third term represents the Zeeman's energy of the decorating spins $\mu$.

At zero temperature, the system passes from the antiferromagnetically ordered ground state to the paramagnetic one when the magnetic field applied on decorating spins exceeds the critical value $h_c=J$. The former ground state is characterized by a perfect antiferromagnetic arrangement of decorating spins placed on horizontal and vertical bonds, while in the latter ground state the system is broken into a set of $2N$ spins polarized towards the magnetic-field direction due to a frustration of the nodal spins $\sigma$. At finite temperatures, the existence of the  antiferromagnetic long-range order terminates at the critical temperature~$T_c$ of the second-order phase transition, which monotonously decreases with increasing the magnetic field until it entirely tends to zero at $h_c=J$.

\section{MAGNETOCALORIC PROPERTIES}
\label{results}

Since the Fisher's super-exchange model~(\ref{eq:H}) is exactly solvable within the decoration-iteration mapping transformation (for more computational details see works~\mbox{\cite{4,5}}), it provides an excellent paradigmatic example of an exactly soluble two-dimensional spin system, which allows an examination of the MCE in a vicinity of the continuous (second-order) phase transition at non-zero magnetic fields. Actually, the magnetocaloric quantities, such as the isothermal entropy change $\Delta S_{T}$ and the adiabatic temperature change $\Delta T_{ad}$ upon the magnetic-field variation $\Delta h\!\!: 0\to h$ can be rigorously calculated by using the formulas:
\begin{eqnarray}
\label{eq:dS}
\Delta S_{T}(T, \Delta h)&=& S(T, h\!\neq0) - S(T, h\!=0),\\[1mm]
\label{eq:dT}
\Delta T_{ad}(S, \Delta h)&=& T(S, h\!\neq0) - T(S, h\!=0).
\end{eqnarray}
Recall that the former relation~(\ref{eq:dS}) is valid if the temperature $T$ of the model is constant, while the latter one~(\ref{eq:dT}) satisfies the adiabatic condition \mbox{$S(T,   h\!\neq0) = S(T, h\!=0)$}.

\begin{figure}[ht!]
\begin{center}
\vspace{-0.0cm}
\includegraphics[angle = 0, width = 1.0\columnwidth]{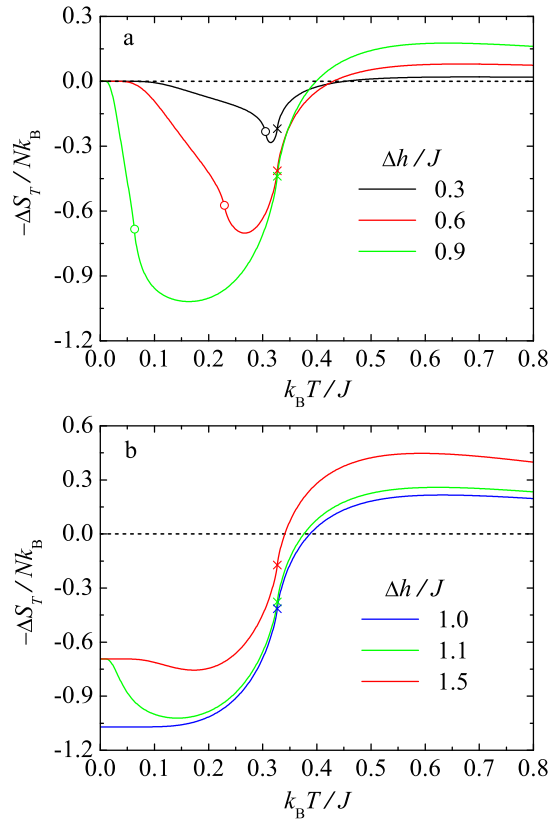}
\vspace{-0.9cm}
\caption{\small Isothermal entropy change normalized per site of the original square lattice versus temperature for several fixed values of the magnetic-field change.}
\label{fig1}
\end{center}
\vspace{-0.45cm}
\end{figure}

Figure~\ref{fig1} illustrates temperature dependencies of the isothermal entropy change ($-\Delta S_{T}$) normalized per site of the original square lattice for various values of the magnetic-field change $\Delta h\!\!: 0\to h$. Crosses on relevant curves mark weak singularities of the zero-field entropy found at the critical temperature $k_{\rm B}T_c/J \simeq 0.3271$, while open circles denote weak singularities of the entropy located at critical points of second-order phase transitions at finite magnetic fields $h/J = 0.3, 0.6$ and $0.9$. As one can see, the magnetocaloric potential $-\Delta S_{T}$ may be either positive or negative  depending on the temperature, which points to both conventional ($-\Delta S_{T}>0$) and inverse ($-\Delta S_{T}<0$) MCE for any value of $\Delta h$. In the high-temperature region $T\gg T_c$, where only short-range ordering occurs, $-\Delta S_{T}$ slowly increases to the broad maximum with decreasing temperature due to suppression of a spin disorder by the applied magnetic field. At certain temperature, $-\Delta S_{T}$ starts to decrease and changes sign from positive to negative as $T$ further decreases.
The zero-temperature limits of $-\Delta S_{T}$ are consistent with the entropy values $S/Nk_{\rm B} = 0$, $0.6931$ and $1.0705$ of the antiferromagnetically ordered ground state, the paramagnetic ground state and the coexistence of both phases at a first-order phase transition, respectively (for more details see Ref.~\cite{5}). Minima in low-temperature parts of $-\Delta S_{T}(T)$ curves observed around the temperature interval \mbox{$T_c(h\!\neq0) < T < T_c(h\!=0)$} for magnetic-field changes $\Delta h\in\left(0, J\right)$ clearly indicate a large inverse MCE slightly above the second-order phase transition (see Fig.~\ref{fig1}a). The origin of this phenomenon can be attributed to strong thermal fluctuations of spins leading to an unusual thermally-induced increase of total magnetization in this region (compare $-\Delta S_{T}(T)$ curves plotted in Fig.~\ref{fig1}a with thermal variations of total magnetization shown in Fig.~9 of Ref.~\cite{4}). In accordance with this statement, the inverse MCE gradually increases and shifts to lower temperatures upon the increase of the field change $\Delta h$. It is obvious from Fig.~\ref{fig1} that $\Delta S_{T}(T,\Delta h=J)<\Delta S_{T}(T,\Delta h\neq J)$ is always satisfied. Thus, one may conclude that the most pronounced inverse MCE can be found for $\Delta h = J$, which exactly coincides with the critical field $h_c = J$ of the first-order phase transition between the magnetically ordered and paramagnetic ground states. If $h > J$, the inverse MCE (minimum in $-\Delta S_{T}(T)$ curves) is gradually reduced with the increasing $\Delta h$ due to weakening of thermal excitations from paramagnetic ground state towards the antiferromagnetically ordered excited state (see Fig.~\ref{fig1}b).

\begin{figure}[ht!]
\begin{center}
\vspace{-0.0cm}
\includegraphics[angle = 0, width = 1.0\columnwidth]{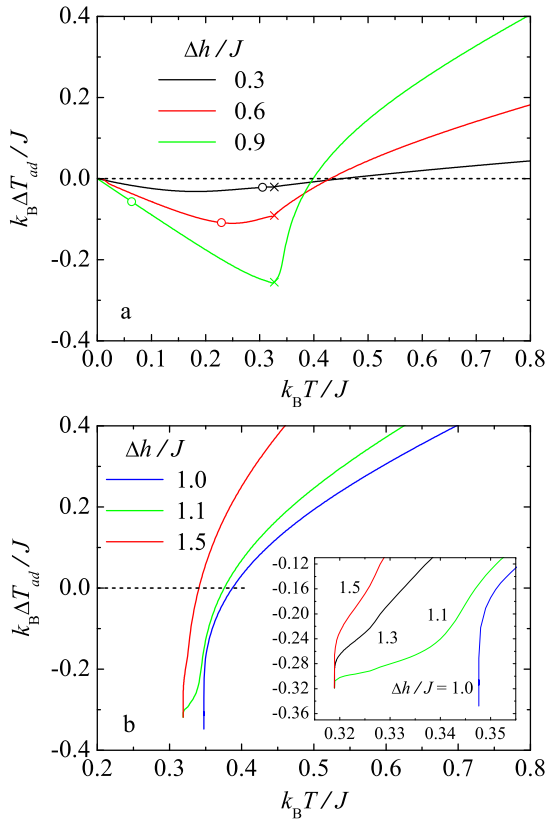}
\vspace{-0.9cm}
\caption{\small Adiabatic temperature change versus temperature for several fixed values of the magnetic-field change.}
\label{fig2}
\end{center}
\vspace{-0.45cm}
\end{figure}

To discuss the MCE, one may alternatively investigate the adiabatic temperature change $\Delta T_{ad}$ of the system at various magnetic-field changes $\Delta h\!\!: 0\to h$. Temperature variations of this magnetocaloric potential for the considered spin model are displayed in Fig.~\ref{fig2}. All curves plotted in Fig.~\ref{fig2} were calculated using Eq.~(\ref{eq:dT}) by keeping the entropy constant. Crosses and open circles on relevant curves determine positions of critical temperatures for $h = 0$ and $h\in\left(0, J\right)$, respectively. Obviously, the adiabatic temperature change $\Delta T_{ad}$ clearly allows to distinguish the conventional MCE ($\Delta T_{ad}>0$) from the inverse MCE ($\Delta T_{ad}<0$). In accordance to the previous discussion, the investigated model heats up faster in a vicinity of the first-order phase boundary between antiferromagnetically ordered ground state and the paramagnetic ground state achieved upon the adiabatic reduction of the magnetic field. Indeed, the magnitude of the negative peak in $\Delta T_{ad}(T)$ curves gradually increases with magnetic-field change and shifts towards the zero-field critical temperature $k_{\rm B}T_c/J \simeq 0.3271$ as the applied magnetic field approaches the critical value $h_c = J$ (see Fig.~\ref{fig2}a). In addition, $\Delta T_{ad}$ versus temperature plot ends at zero value in the asymptotic limit of zero temperature for any $\Delta h\in\left(0, J\right)$, which can be attributed to a perfect antiferromagnetic order of decorating spins placed on horizontal and vertical bonds of the square lattice at zero temperature.
By contrast, the adiabatic temperature change rapidly drops to the finite values $k_{\rm B}\Delta T_{ad}/J = -0.3477$ and $-0.3189$ at the temperatures $k_{\rm B}T/J = 0.3477$ and $0.3189$ for the magnetic-field changes $\Delta h \geq J$ (see Fig.~\ref{fig2}b). In this particular case, the magnetocaloric potential $\Delta T_{ad}$ cannot be defined below aforementioned temperatures, because there is no temperature end point in the adiabatic process if $\Delta h \geq J$. This intriguing behaviour is evidently caused by residual entropies $S/Nk_{\rm B} = 0.6931$ and $1.0705$ detected within the paramagnetic ground state and the coexistence point at the first-order phase transition $h_c = J$, respectively.

\section{Summary}
\label{summary}
In this paper, we have investigated magnetocaloric properties of the exactly solved spin-1/2 Fisher's super-exchange antiferromagnet by using the known exact solution for the magnetic entropy of the model~\cite{5}. The temperature dependencies of the isothermal entropy change and the adiabatic temperature change have been particularly examined for various values of the magnetic-field change. This study has enabled us to clarify the magnetic refrigeration efficiency of the model in a vicinity of the critical temperature of the second-order phase transition, which completely destroys the antiferromagnetic long-range order. The obtained results for both magnetocaloric potentials clearly indicate on the fast heating of investigated spin system during the adiabatic demagnetization process (on a presence of the enhanced inverse MCE) in this region due to strong thermal spin fluctuations leading to the thermally-induced increase of the total magnetization. The maximal heating efficiency of the system has been observed for the magnetic-field change $\Delta h = J$, which coincides with the critical field $h_c = J$ of the first-order  phase transition between the antiferromagnetically ordered and paramagnetic phases.
\\[4mm]
{\bf Acknowledgments}:
This work was financially supported by Ministry of Education, Science,
Research and Sport of the Slovak Republic provided under the VEGA grant No. 1/0043/16 and by  the grant Slovak Research and Development Agency provided under the contract No.~APVV-0097-12.

\end{document}